\documentclass [superscriptaddress,floatfix,showpacs,preprint] {revtex4}

\usepackage{amsmath,amssymb,epsfig,psfig,color,pstricks}

\renewcommand{\red}{\black}

\begin{document}

\title{Comparative study of correlation effects in CaVO$_3$ and SrVO$_3$}

\author{I.A.~Nekrasov}
\affiliation{Institute of Metal Physics, Russian Academy of Sciences-Ural Division,
620219 Yekaterinburg GSP-170, Russia}
\affiliation{Theoretical Physics III, Center for Electronic Correlations and Magnetism,
University of Augsburg, 86135 Augsburg, Germany}
\author{G.~Keller}
\affiliation{Theoretical Physics III, Center for Electronic Correlations and Magnetism,
University of Augsburg, 86135 Augsburg, Germany}
\author{D.E.~Kondakov}
\affiliation{Institute of Metal Physics, Russian Academy of Sciences-Ural Division,
620219 Yekaterinburg GSP-170, Russia}
\affiliation{Theoretical Physics III, Center for Electronic Correlations and Magnetism,
University of Augsburg, 86135 Augsburg, Germany}
\affiliation{Department of Theoretical Physics and Applied Mathematics, USTU, 620002
Yekaterinburg Mira 19, Russia}
\author{A.V.~Kozhevnikov}
\affiliation{Institute of Metal Physics, Russian Academy of Sciences-Ural Division,
620219 Yekaterinburg GSP-170, Russia}
\affiliation{Theoretical Physics III, Center for Electronic Correlations and Magnetism,
University of Augsburg, 86135 Augsburg, Germany}
\affiliation{Department of Theoretical Physics and Applied Mathematics, USTU, 620002
Yekaterinburg Mira 19, Russia}
\author{Th.~Pruschke$^\ast$}
\affiliation{Theoretical Physics III, Center for Electronic Correlations and Magnetism,
University of Augsburg, 86135 Augsburg, Germany}
\author{K.~Held}
\affiliation{Max Planck Institute for Solid State Research, Heisenbergstr.~1, 70569
Stuttgart, Germany}
\author{D.~Vollhardt}
\affiliation{Theoretical Physics III, Center for Electronic Correlations and Magnetism,
University of Augsburg, 86135 Augsburg, Germany}
\author{V.I.~Anisimov}
\affiliation{Institute of Metal Physics, Russian Academy of Sciences-Ural Division,
620219 Yekaterinburg GSP-170, Russia}
\affiliation{Theoretical Physics III, Center for Electronic Correlations and Magnetism,
University of Augsburg, 86135 Augsburg, Germany}

\begin{abstract}
We present parameter-free LDA+DMFT (local density approximation +
dynamical mean field theory) results for the many-body spectra
 of cubic SrVO$_3$ and orthorhombic CaVO$_3$.
Both systems are found to be strongly correlated metals, but {\em
not} on the verge of a metal-insulator transition. 
In spite of the considerably smaller V-O-V bond angle in
CaVO$_{3}$ the LDA+DMFT spectra of the two systems for energies
$E<E_F$ are very similar, their quasiparticle parts being almost
identical. The calculated spectrum for $E>E_F$ shows more
pronounced, albeit still small, differences. This is in contrast
to earlier theoretical and experimental conclusions, but in good
agreement with recent bulk-sensitive photoemission and x-ray
absorption experiments.
\end{abstract}
\pacs{71.27.+a, 71.30.+h}
\vspace{-0.5cm}
\maketitle

\section{Introduction}

Transition metal oxides are an ideal laboratory for the study of electronic
correlations in solids. In these materials, the $3d$ bands are comparatively
narrow with width $W\approx~2-3\,$~eV so that electronic
correlations, induced by the local Coulomb interaction
$\bar{U}\approx~3-5\,$~eV, are strong.
For $\bar{U}/W\ll~1$,
 we have a weakly correlated metal and
the local density approximation (LDA) (see~\cite{JonGun} for a
review) works. In the opposite limit $\bar{U}/W\gg1$ and for an
integer number of $3d$ electrons, we have a Mott insulator with
two separate Hubbard bands as described by Hubbard's I and III
approximations~\cite{Hubbard} or the LDA+U method~\cite{LDAU}.
Transition metals are, however, in the ``in-between'' regime,
$\bar{U}/W={\cal O}(1)$, where the metallic phase is strongly
correlated with coexisting {\em quasiparticle} peak at the Fermi
level and Hubbard side bands.

Dynamical mean-field theory (DMFT)~\cite{MetzVoll89,vollha93,pruschke,georges96,PT},
a modern, non-perturbative many-body approach,
is able to capture the physics in the whole
range of parameters from $\bar{U}/W\ll 1$ to  $\bar{U}/W\gg1$
for model Hamiltonians, like the one-band Hubbard model (for an
introduction into DMFT and its applications see~\cite{PT}).
And, with the recent merger~\cite{poter97,Held01,licht} of LDA and DMFT,
we are now able to handle this kind of correlation physics
within realistic calculations for transition metal oxides.

A particularly simple transition metal oxide system is Ca$_{1-x}$Sr$_x$VO$_3$
with a $3d^1$ electronic configuration and a cubic perovskite
lattice structure, which is ideal for SrVO$_3$ and
orthorhombically distorted upon increasing the Ca-doping $x$.
Fujimori {\it et al.}~\cite{Fujimori92a}
initiated the present interest in this  3$d^{1}$ series,
reporting
a pronounced lower Hubbard band in the photoemission spectra (PES)  which
cannot be explained by conventional LDA.
Subsequently, thermodynamic properties were studied,
and the Sommerfeld coefficient,
resistivity, and paramagnetic susceptibility were found to
be essentially independent of $x$~\cite{old_experiments}.

On the other hand, PES~\cite{Morikawa95},
 seemed to imply dramatic differences
between  CaVO$_{3}$ and SrVO$_{3}$, with CaVO$_{3}$
having (almost) no spectral weight at the Fermi energy.
 Rozenberg {\it et al.}~\cite{Rozenberg96}
interpreted these results in terms of
 the (DMFT) Mott-Hubbard transition
in the half-filled
single-band Hubbard model. With the same
theoretical ansatz, albeit different parameters,
the optical
conductivity~\cite{Makino98} and the PES  experiments~\cite{Maiti01}
for different dopings $x$  were reproduced.

The puzzling discrepancy between these spectroscopic and the thermodynamic
measurements, suggesting a Mott-Hubbard metal-insulator transition
or not,
were resolved recently by
bulk-sensitive PES obtained by Maiti~\textit{et al.}~\cite{Maiti01} and
Sekiyama~\textit{et al.}~\cite{Sekiyama02,Sekiyama03}.
 In the latter work it was shown
that (i) the technique of preparing the sample surface (which should preferably
be done by fracturing) is very important, and that (ii) the energy of the X-ray
beam should be large enough to increase the photoelectron escape depth to
achieve bulk-sensitivity. Theoretically, pronounced differences between
  SrVO$_3$ surface and bulk spectra were reported by Liebsch
\cite{Liebsch02}. Besides bulk-sensitivity, the beam should provide a high instrumental
resolution (about 100~meV in~\cite{Sekiyama02,Sekiyama03}).
 With these experimental
improvements, the PES of SrVO$_{3}$ and CaVO$_{3}$ were found to
be almost identical~\cite{Maiti01,Sekiyama02,Sekiyama03}, implying
consistency of spectroscopic and thermodynamic results at last.
This is also in accord with earlier 1s x-ray absorption spectra
(XAS) by Inoue {\em et al.}~\cite{Inoue94} which differ only
slightly  above the Fermi energy, as well as with BIS data
\cite{Morikawa95}. Parameterfree LDA+DMFT results by us
\cite{HISTORY}, reported in the joint experimental and theoretical
Letter by Sekiyama {\em et al.}~\cite{Sekiyama03}, showed good
agreement with the high-resolution bulk-sensitive PES. Independent
LDA+DMFT calculations for 3d$^1$ vanadates, including SrVO$_3$ and
CaVO$_3$ have been reported by Pavarini {\em et
al.}~\cite{Pavarini03}. Most recently Anisimov {\em et al.}
employed a full-orbital computational scheme based on Wannier
functions formalism for LDA+DMFT to calculate \emph{total}
densities of states for SrVO$_3$ which are in good agreement with
new PES data. \cite{Wannier}

In this paper we present details of our LDA+DMFT calculation
\cite{HISTORY} which could not be included in the joint Letter
\cite{Sekiyama03}. In particular, we discuss the crystal structure
of SrVO$_3$ and CaVO$_3$, explain the origin of the surprisingly
small change in bandwidth of these two systems, and compare the
calculated spectrum for energies $E>E_F$ with XAS~data
\cite{Inoue94}.
We start with discussing the structural differences between
SrVO$_3$ and CaVO$_3$  in Section~\ref{Struct}, based on the most
recent crystallographic data provided by Inoue~\cite{Inoue03}. How
these differences reflect in the LDA results is shown in
Section~\ref{lda_res}, before we turn to the LDA+DMFT scheme in
Sections~\ref{dmft} and the effects of electronic correlations in
Section \ref{dmft_res}. Section~\ref{conc} concludes our
presentation.

\section{SrVO$_3$ vs. CaVO$_3$: Structural differences and LDA results}
\subsection{Structural differences}
\label{Struct}

\begin{figure}
\centering
\epsfig{file=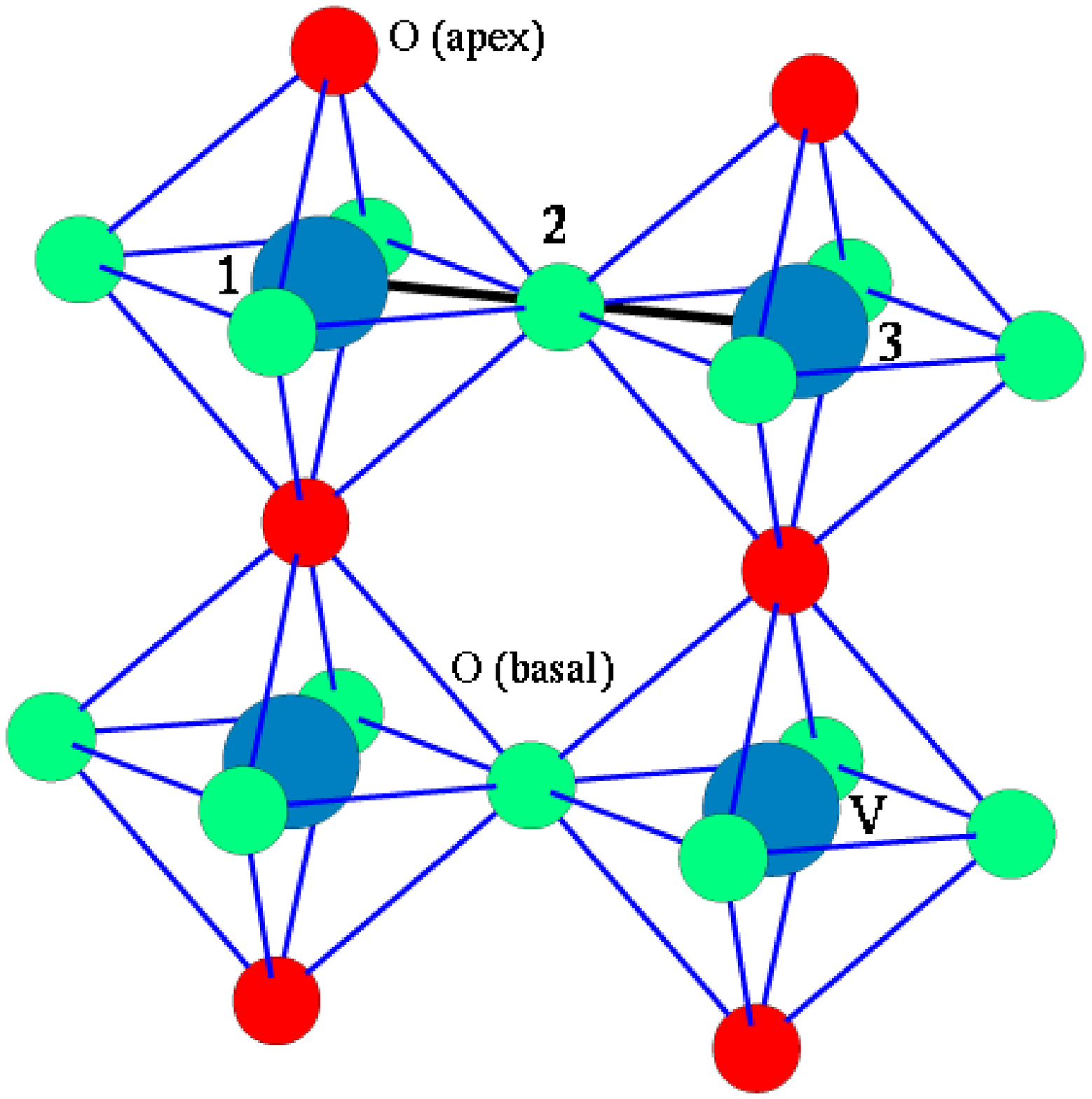,width=0.5\textwidth}
\caption{SrVO$_3$, an ideal cubic perovskite, with
 V-O-V angle $\theta$=$\angle$123=180$^\circ$; V: large ions,
O$_{\rm basal}$: small bright ions, O$_{\rm apex}$: small dark
ions.}
\label{Sr_crystal}
\end{figure}

\begin{figure}
\centering
\epsfig{file=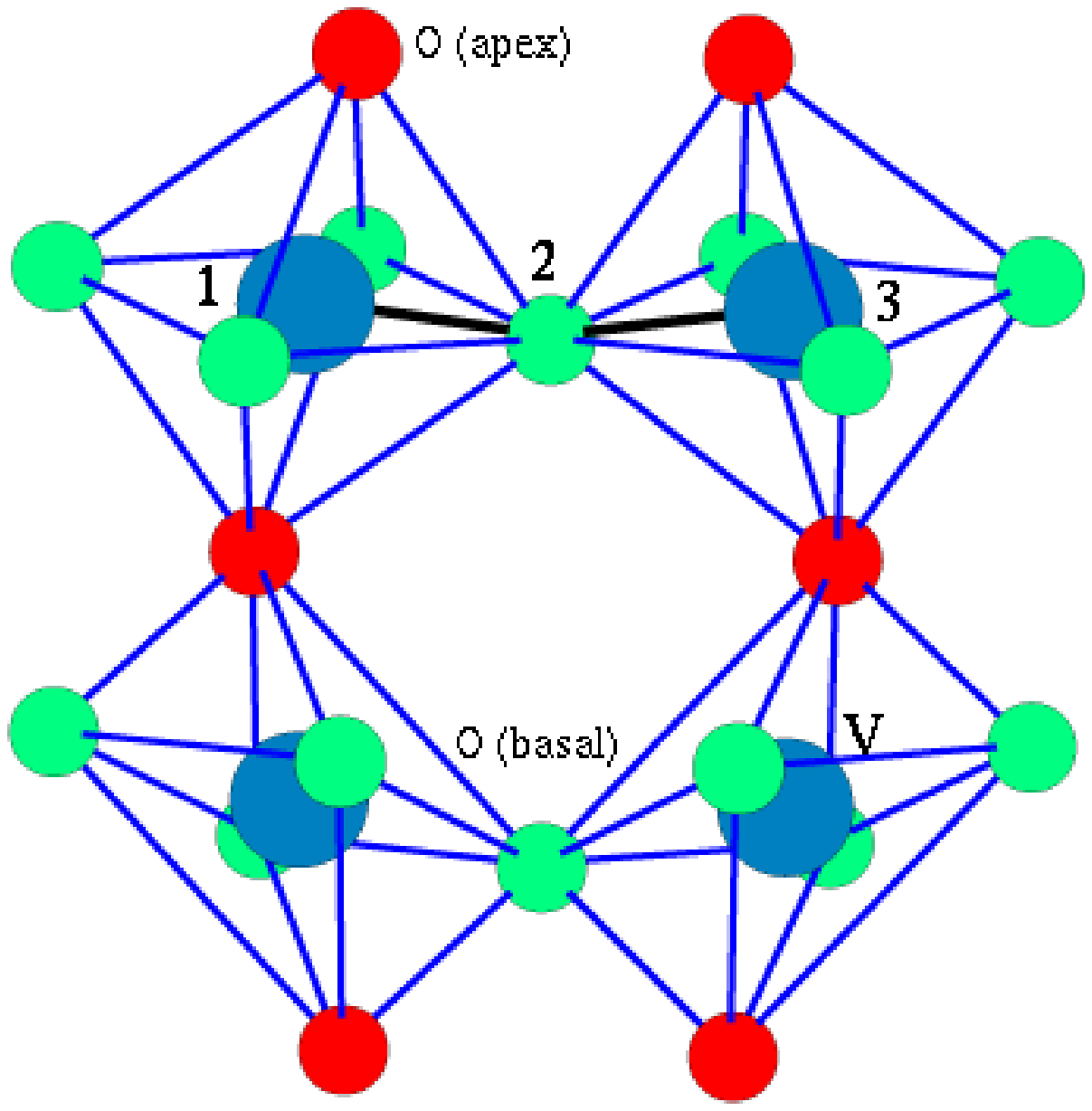,width=0.5\textwidth}
\caption{CaVO$_3$, an orthorhombicaly distorted perovskite, with
V-O-V angles $\theta_{basal}\approx161^\circ$, $\theta_{plane}\approx163^\circ$;
V: large ions,
O$_{\rm basal}$: small bright ions, O$_{\rm apex}$: small dark
ions. The local c-axis, later used for the definition of the orbitals, is directed along the tilted
 V-O$_{\rm apex}$ direction.}
\label{Ca_crystal}
\end{figure}

Both SrVO$_3$ and CaVO$_3$ are  perovskites, with
SrVO$_3$ in the ideal cubic structure Pm$\bar 3$m~\cite{Rey90} (Fig.\ \ref{Sr_crystal}) and CaVO$_3$ in the  orthorhombicaly distorted (GdFeO$_3$) Pbnm
structure~\cite{Chamberland71,Inoue03} (Fig.\ \ref{Ca_crystal}).
The ideal cubic structure of SrVO$_3$ implies that the VO$_6$ octahedra,
the main structural elements of the crystal,
are not distorted: The V-O$_{\rm basal}$ and V-O$_{\rm apex}$ distances are the same,
the
O$_{\rm apex}$-V-O$_{\rm basal}$ and O$_{\rm basal}$-V-O$_{\rm basal}$ angles are 90$^\circ$,
and the V-O-V bond angles are exactly 180$^\circ$.
The substitution of the large Sr$^{2+}$ ions by
smaller Ca$^{2+}$ ions leads to a tilting, rotation, and distortion of the
VO$_6$
octahedra.
Nonetheless, in CaVO$_3$, both the V-O$_{\rm basal}$ and the V-O$_{\rm apex}$ distances are
practically the same. Therefore, the most important feature of the distortion
is the change of the V-O$_{\rm basal}$-V and V-O$_{\rm apex}$-V angles,
which determines the strength of the effective V3$d$-V3$d$ hybridization and, hence,
the bandwidth.
Due to the rotation and
tilting of the octahedra both angles (one of them is marked as $\angle$123 in
Figs.~\ref{Sr_crystal} and \ref{Ca_crystal}) have the
same value $\theta = 180^\circ$ for SrVO$_3$ and
$\theta_{basal}\approx161^\circ$, $\theta_{plane}\approx163^\circ$for CaVO$_3$.

\subsection{LDA results}
\label{lda_res}

\begin{figure}
\centering
\epsfig{file=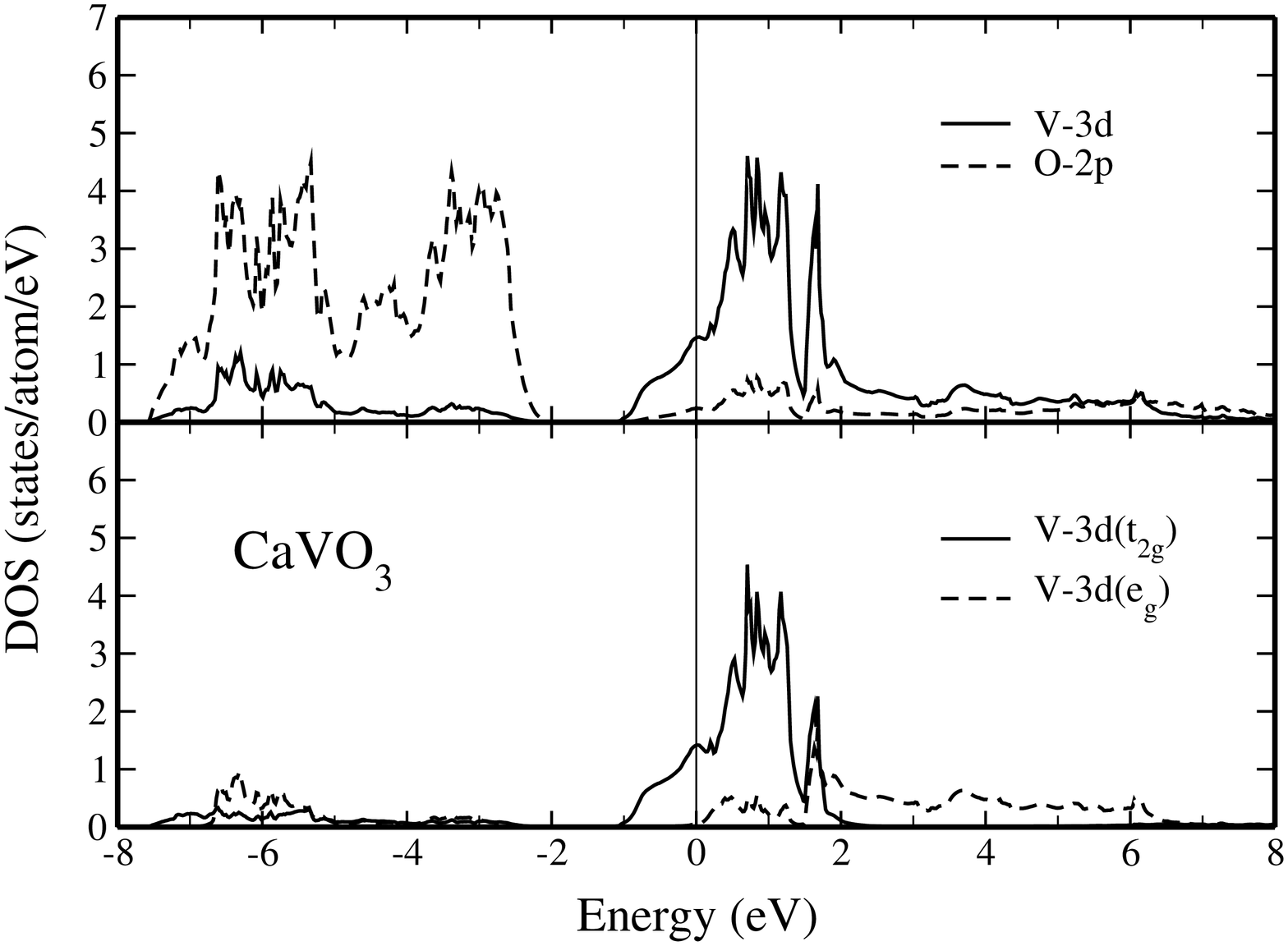,angle=270,width=0.7\textwidth}
\caption{DOS of CaVO$_3$ as
calculated by LDA(LMTO). Upper panel:  V-3d
(full line) and O-2$p$ (dashed line) DOS; lower panel:
partial DOS of V-3$d$ ($t_{2g}$) (full line) and V-3d($e_{g}$)
(dashed line) orbitals. The Fermi level corresponds to 0$\,$eV.}
\label{fig_LDA1b}
\end{figure}

The valence states of SrVO$_3$ and CaVO$_3$ consist of completely occupied
oxygen 2$p$ bands and partially occupied V 3$d$ bands.
Since the V ion has a formal oxidation of 4+,
there is one electron  per V ion
occupying the $d$-states (configuration $d^1$).
In Figs.~\ref{fig_LDA1a} and~\ref{fig_LDA1b} the LDA density of states (DOS)
 are presented
for both compounds, as calculated by the
linearized muffin-tin orbitals method (LMTO)~\cite{LMTO} with an orthogonal
 basis set of
Sr($5s,5p,4d,4f$), Ca($4s,4p,3d$), V($4s,4p,3d$), and O($3s,2p,3d$) orbitals~\cite{basis}.
These results are in agreement with previous LDA calculations in the
basis of augmented plane waves (APW) by Tagekahara \cite{Tagekahara}.
For a combination of the LDA band structure with
many-body techniques~\cite{poter97} however,
the atomic-like LMTO wave functions employed here are
particularly convenient.
As is apparent from Figs.\ \ref{fig_LDA1a} and \ref{fig_LDA1b},  the O-2$p$
states lie between -7.5eV and -2.0eV in SrVO$_3$ and between -6.5eV and -1.5eV
in CaVO$_3$. The 3$d$ states of V are located between -1.1eV and
6.5eV in SrVO$_3$ and between -1.0eV and 6.5eV in CaVO$_3$.
These results are in agreement with previous LDA calculations in the
basis of augmented plane waves (APW) by Tagekahara \cite{Tagekahara},
but for our
purposes LMTO   is more appropriate.

In both compounds, the V ions are in a octahedral coordination of oxygen ions.
The octahedral crystal field splits the V-3$d$  states into three
t$_{2g}$ orbitals and two e$_g$ orbitals. In the cubic symmetry
of SrVO$_3$,
hybridization  between t$_{2g}$ and  e$_g$ states is forbidden,
and the orbitals within both subbands are degenerate. In contrast,
the distorted orthorhombic structure of CaVO$_3$ allows them to mix
and the degeneracy is lifted.
In the lower part of Figs.\ \ref{fig_LDA1a} and \ref{fig_LDA1b},
 we present these t$_{2g}$ and e$_g$ subbands of the V-3$d$ band.
For both Sr and Ca compounds,
the reader can see an admixture of these Vanadium t$_{2g}$ and  e$_g$
states to the oxygen 2$p$ states in the energy region [-8$\,$eV,-2$\,$eV].
This is due to hybridization and amounts  to
 12\% and 15\%  of the  total t$_{2g}$ weight for SrVO$_3$ and CaVO$_3$,
respectively.
The main t$_{2g}$ (e$_g$) weight lies in the energy region
[-1.1$\,$eV,1.5$\,$eV] ([0.0$\,$eV,6.5$\,$eV]) in SrVO$_3$ and
[-1.0$\,$eV,1.5$\,$eV] ([0.2$\,$eV,6.5$\,$eV]) in CaVO$_3$. Since the energy difference
between the centers of gravities of the t$_{2g}$ and  e$_g$ subbands
is comparable with the bandwidth, t$_{2g}$ and  e$_g$ states can be considered
as sufficiently separated in both compounds.

Let us now concentrate on the  t$_{2g}$ orbitals crossing the
Fermi energy and
compare the LDA t$_{2g}$ DOS of SrVO$_{3}$ and CaVO$_{3}$. Most importantly,
the one-electron t$_{2g}$ LDA bandwidth  of CaVO$_{3}$, defined as the energy
interval where the t$_{2g}$ DOS in Figs.\ \ref{fig_LDA1a} and
\ref{fig_LDA1b} is non-zero, is found to be only
4\% smaller than that of SrVO$_{3}$ ($W_{\mathrm{CaVO_{3}}}=2.5$~eV,
$W_{\mathrm{SrVO_{3}}}=2.6$~eV);
in general agreement with other LDA calculations \cite{Pavarini03}.
This is in contrast to the expectation that the strong
lattice distortion with a decrease of the V-O-V bond angle from 180$^{\circ }$
to 162$^{\circ }$ affects the t$_{2g}$ bandwidth much more strongly. Such a
larger effect indeed occurs in the  e$_{g}$ bands for which the
 bandwidth  is reduced
by 10\%. To physically understand the smallness of the  narrowing  of the
t$_{2g}$ bands we have investigated a larger Hamiltonian consisting of e$_{g}$,
t$_{2g}$, and oxygen $p$ orbitals.
We calculated the overall direct $d\!\!-\!\!d$ and the indirect 
$d\!\!-\!\!p\!\!-\!\!d$ hopping for this larger
Hamiltonian. The predominant contribution to the
e$_{g}$-e$_{g}$ hopping is  through a $d\!\!-\!\!p\!\!-\!\!d$ hybridization,
which is considerably decreasing with the lattice distortion. This is also the
case for the $t_{2g}$ orbitals. However, for the  $t_{2g}$ orbitals, the  direct
$d\!\!-\!\!d$ hybridization is also important. This hybridization
{\em increases} with the distortion since the  $t_{2g}$ orbital lobes point more
directly towards each other
in the distorted structure.  
Hopping parameters for individual orbitals change rather
substantially from SrVO$_3$ to CaVO$_3$.
But, altogether, a decreasing $d\!\!-\!\!p\!\!-\!\!d$ hybridization and
an increasing  $d\!\!-\!\!d$ hybridization results in a very small change of
the t$_{2g}$ bandwidth.

\section{Correlation effects in SrVO$_3$ and CaVO$_3$}
\subsection{Dynamical Mean-Field Theory}
\label{dmft}

As is well-known, LDA does not treat the effects of strong local Coulomb
correlations adequately. To overcome this drawback, we  use LDA+DMFT as a
non-perturbative approach to study strongly correlated systems~\cite{poter97,Held01,licht}.
It combines the strength of the LDA in describing weakly
correlated electrons in the $s$- and $p$-orbitals, with the DMFT treatment
of the dynamics due to local Coulomb interactions. In the present paper we
will discuss the relevant parts of the LDA+DMFT approach only briefly,
refering the reader to Ref.~\cite{Held01,licht} for details.

 The LDA+DMFT  Hamiltonian
can be written as
\begin{equation}
\hat{H} =\hat{H}_{\rm LDA}^{0}+U\sum_{m}\sum_{i}
\hat{n}_{im\uparrow}\hat{n}_{im\downarrow } +
\;\sum_{i\;m\neq m'\;\sigma \sigma'}\;
(U'-\delta_{\sigma \sigma'}J)\;
\hat{n}_{im\sigma}\hat{n}_{im'\sigma'}.
\label{H}
\end{equation}
Here, the index $i$ enumerates the V sites, $m$ the individual t$_{2g}$  orbitals and
$\sigma$ the spin.
$H_{{\rm LDA}}^{0}$ is the one-particle Hamiltonian generated from the LDA band
structure with an averaged Coulomb interaction subtracted to account for
double counting~\cite{poter97}; $U$ and $U^\prime$ denote
 the local intra-orbital and inter-orbital Coulomb repulsions,
and $J$ is the exchange interaction.

 We calculated
these interaction strengths by means of the constrained LDA
method~\cite{Gunnarsson89} for SrVO$_3$, allowing the e$_{g}$
states to participate in screening~\cite{Solovyev96}. The
resulting value of the averaged Coulomb interaction is
$\bar{U}=3.55$~eV ($\bar{U}=U^{\prime }$ for t${_{2g}}$
orbitals~\cite{Held01,Zoelfl00}) and $J=1.0$~eV. The intra-orbital
Coulomb repulsion $U$ is then fixed by rotational invariance to
$U=U^{\prime }+2J = 5.55$~eV.
We did not calculate $\bar{U}$ for CaVO$_{3}$ because the standard procedure
to calculate the  Coulomb interaction parameter between two t${_{2g}}$
electrons,
 screened by e$_{g}$ states, is not applicable
for the distorted crystal structure where the e$_{g}$ and t$_{2g}$ orbitals
are not separated by symmetry. On the other hand,
it is well-known that the change of the {\em local}
Coulomb interaction is  typically much smaller than the change
in the DOS, which  was found to depend only very weakly on the bond angle.
That means that  $\bar{U}$ for CaVO$_{3}$ should be nearly
the same as for SrVO$_{3}$. Therefore we used
$\bar{U}=3.55$~eV and $J=1.0$~eV for both SrVO$_{3}$ and
CaVO$_{3}$. This is also in agreement with previous calculations
of vanadium systems~\cite{Solovyev96} and experiments
\cite{Makino98}.

DMFT maps the lattice problem Eq.\
(\ref{H}) onto a self-consistent auxiliary impurity problem,
which is here solved numerically by the quantum Monte-Carlo (QMC)
technique \cite{QMC}.
Combined with the maximum entropy method~\cite{MEM},
this technique allows us to calculate spectral functions
\cite{Liebsch00,Nekrasov00,Held01a,Nekrasov02}.

A computationally important simplification is due to the fact that,
in ideal cubic perovskites, the (degenerate) t$_{2g}$ states do not mix
with the (degenerate) e$_g$ states. In this particular case, the self-energy
matrix $\Sigma (z)$ is diagonal with respect to the orbital indices.
Under this condition, the Green functions $G_m(z)$ of the lattice
problem can be expressed in the DMFT self-consistency equation
by the  Hilbert transform of the non-interacting DOS
$N^{(0)}_m(\omega)$:
\begin{equation}
G_m(z)=\int d\epsilon
\frac{N^{0}_m(\epsilon )}{z-\Sigma_m(z)-\epsilon}\;\;.
\label{intg}
\end{equation}
This procedure avoids the rather cumbersome and problematic $k$-integration
over the Brillouin zone by the analytical tetrahedron method~\cite{Lambin84}.
We obtain $N^{(0)}_m(\epsilon)$ for SrVO$_3$ and CaVO$_3$ from
the $t_{2g}$-projected DOS of
Figs.\ \ref{fig_LDA1a} and \ref{fig_LDA1b} by truncating the
contribution in the oxygen range below -1.5$\,$eV and multiplying
by a renormalization factor
(to have maximally one electron per site and orbital).
This procedure is employed as it resembles the
 three {\em effective} bands of $t_{2g}$ character which
 cross the Fermi energy
and are, hence, responsible for the low-energy physics.

\subsection{LDA+DMFT(QMC) results and discussion}
\label{dmft_res}

\begin{figure}
\centering \psfig{file=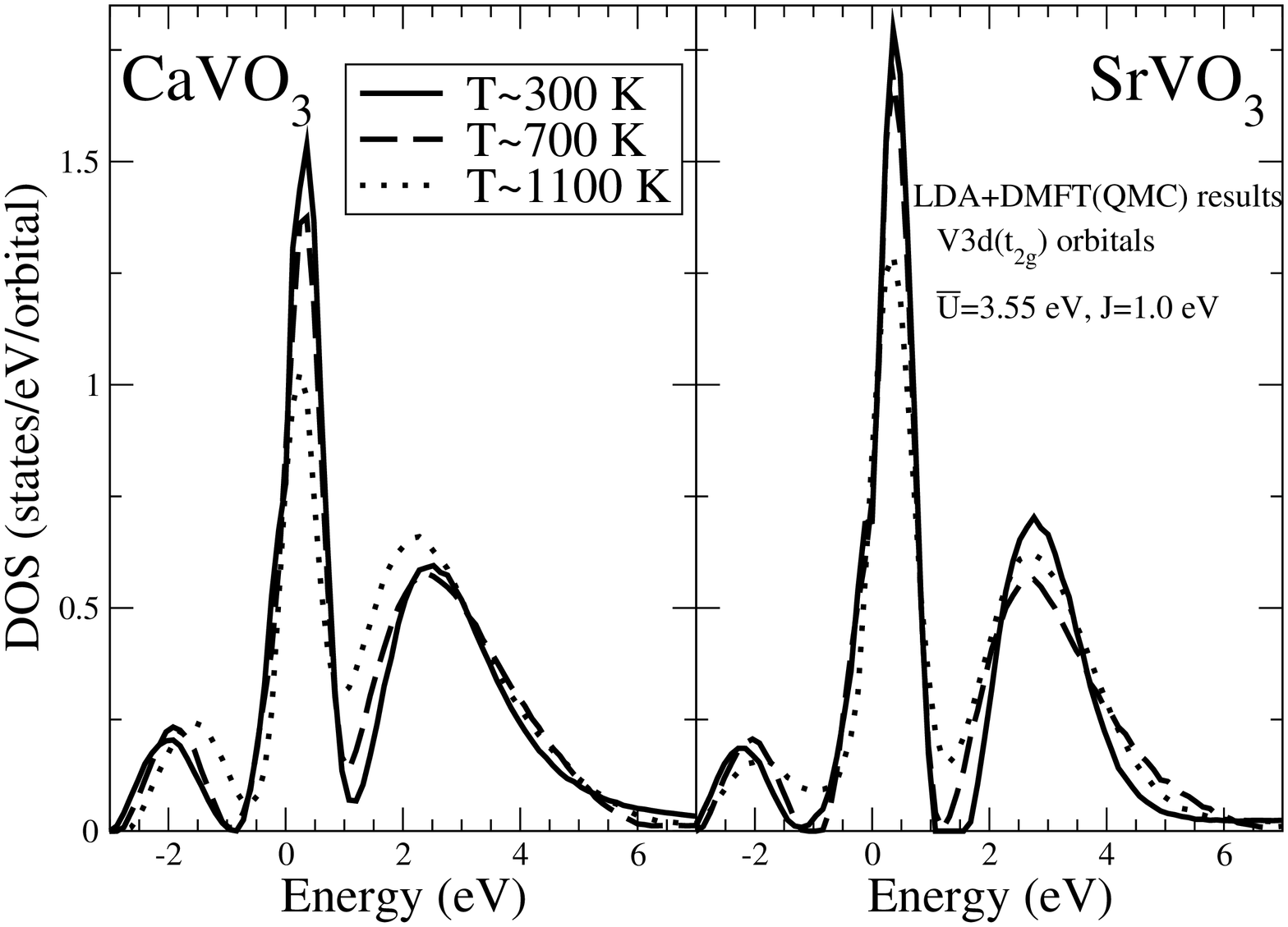,angle=270,width=0.7\textwidth,clip}
\caption{LDA+DMFT(QMC) spectrum of SrVO$_{3}$ (right panel)
and CaVO$_{3}$ (left panel) calculated at T$\approx$300~K
(solid lines), T$\approx$700~K (dashed lines),
and T$\approx$1100~K) (doted lines).}
\label{fig_ldadmft}
\end{figure}

The calculated LDA+DMFT(QMC)
spectra for SrVO$_{3}$ (right panel) and CaVO$_{3}$ (left panel)
are presented in Fig.~\ref{fig_ldadmft} for different temperatures.
Due to genuine correlation effects, a  lower
Hubbard band at about $-1.5$~eV and an upper Hubbard band at about
$2.5$~eV is formed, with a well pronounced quasiparticle peak at the Fermi
energy. The 4\% difference in the LDA bandwidth between
SrVO$_{3}$ and CaVO$_{3}$ is only reflected in some additional
transfer of spectral weight from the quasiparticle peak to the
Hubbard bands, and minor differences in the positions of the
Hubbard bands. Clearly, the two systems are not on the verge of a
Mott-Hubbard metal-insulator transition.  Both, SrVO$_{3}$ and CaVO$_{3}$, are strongly
correlated metals, though SrVO$_{3}$ is slightly
less correlated than CaVO$_{3}$ with
 somewhat more  quasiparticle weight,
in accord with their different
LDA bandwidth. As one can see from Fig.~\ref{fig_ldadmft}, the
effect of temperature on the spectrum is small for $T \lesssim
700$~K.

\begin{figure}
\centering \epsfig{file=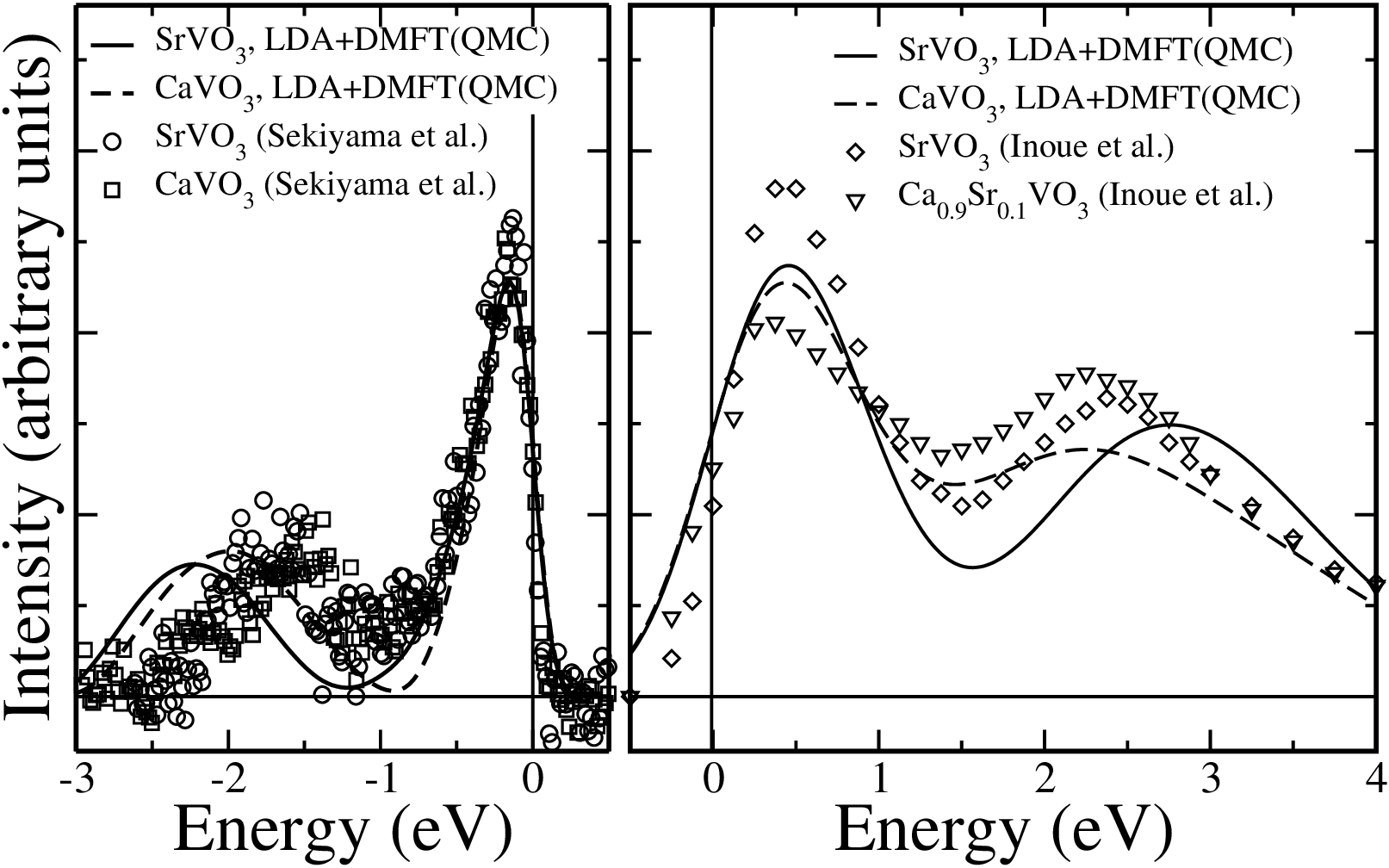,width=0.6\textwidth,clip}
\caption{
{\red
Comparison of the parameter-free LDA+DMFT(QMC) spectra of
SrVO$_{3}$ (solid line) and CaVO$_{3}$ (dashed line)
  with experiments below {\em and} above the Fermi energy.
Left panel: high-resolution PES for SrVO$_{3}$ (circles) and
CaVO$_{3}$ (rectangles)~\protect\cite{Sekiyama02,Sekiyama03}.
Right panel: 1s XAS for SrVO$_{3}$ (diamonds) and
Ca$_{0.9}$Sr$_{0.1}$VO$_{3}$ (triangles)~\protect\cite{Inoue94}.
Horizontal line: experimental subtraction of the background
intensity.} \protect\vspace{-0.3cm} \label{fig_XPS}}
\end{figure}

In the left panel of Fig.~\ref{fig_XPS}, we compare our LDA+DMFT spectra
(300K), which were multiplied with the Fermi function at the experimental
temperature (20$\,$K) and Gauss broadened with the experimental resolution of
$0.1\,$eV \cite{Sekiyama02} with the experimental PES data obtained by
subtracting the experimentally
estimated surface and oxygen contributions.
Due to the high photon energy (several hundred eV) \cite{Sekiyama02},
the PES transition matrix
elements will have {\em no} strong energy dependence in the
energy interval of a few eV studied here, which justifyies
that we do not take such matrix elements into account.

The quasiparticle peaks
in theory and experiment are seen to be in very good agreement. In particular,
their heights and widths are almost identical for both SrVO$_{3}$ and CaVO$_{3}$.
The difference in the positions of the lower Hubbard bands may be partly due to
(i) the subtraction of the (estimated) oxygen contribution which might also
remove some $3d$ spectral weight below $-2$~eV, and (ii) uncertainties in the
{\em{ab-initio}} calculation of $\bar{U}$.

Since inverse photoemission spectroscopy (IPES) data for
SrVO$_{3}$ and CaVO$_{3}$ are not yet available we compare
\cite{IPES} the calculated spectrum for energies $E>E_F$ with XAS
data by Inoue \emph{et al.}~\cite{Inoue94} (right panel of
Fig.~\ref{fig_XPS}). We consider core-hole life time effects by
Lorentz broadening the spectrum with 0.2~eV~\cite{Krause79},
multiply with the inverse Fermi function (80K), and apply Gauss
broadening given by the experimental resolution of
$0.36\,$eV~\cite{Inoue03}. Again, the overall agreement of the
weights and positions of the quasiparticle and upper $t_{2g}$
Hubbard band is good, including the tendencies when going from
SrVO$_{3}$ to CaVO$_{3}$. For CaVO$_{3}$
(Ca$_{0.9}$Sr$_{0.1}$VO$_{3}$ in the experiment), the LDA+DMFT
 quasiparticle weight is somewhat lower than in  experiment.

 In contrast
to  one-band Hubbard model calculations, our material specific results
reproduce the strong asymmetry around the Fermi energy w.r.t. weights and
bandwidths. Our results also give a different interpretation of the XAS than
in~\cite{Inoue94} where the maximum at about $2.5\,$eV was attributed
solely to the $e_g$ band. We find that also the $t_{2g}$ upper Hubbard band
contributes in this energy range.

The slight
differences in the quasiparticle peaks (see Fig.~\ref{fig_ldadmft}) lead to
different effective masses, namely $m^*/m_0\!=\!2.1$ for SrVO$_{3}$ and
$m^*/m_0\!=\!2.4$ for CaVO$_{3}$. These theoretical values agree with $m^{\ast
}/m_{0}\!=\!2-3$ for SrVO$_{3}$ and CaVO$_{3}$ as obtained from de Haas-van
Alphen experiments and thermodynamics \cite{old_experiments,Inoue02}. We note
that the effective mass of CaVO$_{3}$ obtained from optical experiments is
somewhat larger, i.e., $m^{\ast }/m_{0}\!=\!3.9$ \cite{Makino98}.

\section{Conclusion}
\label{conc}

In summary, we investigated the spectral properties of the
correlated 3$d^{1}$ systems SrVO$_{3}$ and CaVO$_{3}$ within the
LDA+DMFT(QMC) approach. Constrained LDA was used to determine the
average Coulomb interaction as $\bar{U}\!=\!3.55$~eV and the
exchange coupling as $J\!=\!1.0$~eV. With this input we calculated
the spectra of the two systems in a parameter-free way. Both
systems are found to be strongly correlated metals, with a
transfer of the majority of the spectral weight from the
quasiparticle peak to the incoherent upper and lower Hubbard
bands. The calculated DMFT spectra of SrVO$_{3}$ and CaVO$_{3}$,
and in particular the quasiparticle parts, are found to be very
similar, differences being slightly more pronounced at energies
above the Fermi energy.
 Our calculated spectra agree very well with recent
bulk-sensitive high-resolution PES \cite{Sekiyama02} and XAS
\cite{Inoue94} data, i.e., with the experimental spectrum below
{\em and} above the Fermi energy. Both compounds are similarly
strongly correlated metals; CaVO$_{3}$  is not on the verge of a
Mott-Hubbard transition.

Our results are in contrast to previous theories and the
expectation that the strong lattice distortion leads to a strong
narrowing of the   CaVO$_{3}$ bandwidth and, hence, to much
stronger correlation effects in CaVO$_{3}$. While the  e$_g$ bands
indeed narrow considerably, the competition between decreasing
$d\!\!-\!\!p\!\!-\!\!d$  and increasing $d\!\!-\!\!d$
hybridization leads to a rather insignificant narrowing of the
t$_{2g}$ bands at the Fermi energy. This explains why the
k-integrated spectral functions of CaVO$_{3}$ and SrVO$_{3}$ are
so similar. With our theoretical results confirming the  new  PES
and XAS experiments, we conclude that the insulating-like behavior
observed in earlier PES and BIS experiments on CaVO$_{3}$ must be
attributed to surface effects~\cite{Liebsch02}.

\section{Acknowledgments}
This work was supported by Russian Basic Research Foundation grant
RFFI-GFEN-03-02-39024\_a (VA,IN,DK), RFFI-04-02-16096 (VA,IN,DK) and
the Deutsche Forschungsgemeinschaft through
Sonderforschungsbereich 484 (DV,GK,IN)  and in
part by the joint UrO-SO project N22 (VA,IN), the Emmy-Noether program (KH), by
Grant of President of Russian Federation for young  scientists MK-95.2003.02 (IN),
Dynasty Foundation and International Center for Fundamental Physics in Moscow
program for young scientists 2004 (IN), Russian Science Support Foundation program
for young PhD of Russian Academy of Science 2004 (IN).

$^\ast$New permanent address: Institut f\"ur Theoretische Physik,
Universit\"at G\"ottingen, Friedrich-Hund-Platz~1,
D-37077 G\"ottingen, Germany.

\pagebreak

\begin{figure}
\centering
\epsfig{file=LDA_DOSes_Sr.eps,angle=270,width=0.6\textwidth}
\caption{DOS of SrVO$_3$ as
calculated by LDA(LMTO). Upper panel:  V-3d
(full line) and O-2$p$ (dashed line) DOS; lower panel:
partial DOS of V-3$d$ ($t_{2g}$) (full line) and V-3d($e_{g}$)
(dashed line) orbitals. The Fermi level corresponds to 0$\,$eV.}
\label{fig_LDA1a}
\end{figure}

\begin{thebibliography}{99}
\bibitem{basis}
Logarithmic derivatives were fixed to the value $-l-1$ for Sr($p,f$), Ca($p$),
V($d$), O($s,d$) states.
\bibitem{IPES} It should be noted that the XAS measurements were
taken at the Oxygen K-edge. Hence there is no symmetry breaking of
the Vanadium 2p shell in the final state. In this case the XAS
data can be taken as a good substitute for presently unavailable
IPES data.
\end{thebibliography}
\end{document}